\documentclass[12pt]{iopart}
\expandafter\let\csname equation*\endcsname\relax
\expandafter\let\csname endequation*\endcsname\relax
\usepackage{amsmath}
\usepackage[T1]{fontenc}
\usepackage{graphicx}
\usepackage{multirow}
\usepackage{xcolor}
\usepackage{xspace}
\usepackage{cite}
\usepackage[colorlinks=true,linkcolor=blue,citecolor=blue,urlcolor=blue]{hyperref}

\newcommand{\Lc}{\ensuremath{\Lambda_{\mathrm{c}}^{+}}\xspace}
\newcommand{\Lb}{\ensuremath{\Lambda_{\mathrm{b}}^{0}}\xspace}
\newcommand{\Dz}{\ensuremath{\mathrm{D}^{0}}\xspace}
\newcommand{\Dstp}{\ensuremath{\mathrm{D}^{*+}}\xspace}
\newcommand{\Dstz}{\ensuremath{\mathrm{D}^{*0}}\xspace}
\newcommand{\Sc}{\ensuremath{\Sigma_{\mathrm{c}}}\xspace}
\newcommand{\Scst}{\ensuremath{\Sigma_{\mathrm{c}}^{*}}\xspace}

\newcommand{\zh}{\ensuremath{z_{h}}\xspace}
\newcommand{\pT}{\ensuremath{p_{\mathrm{T}}}\xspace}
\newcommand{\ptjet}{\ensuremath{p_{\mathrm{T}}^{\mathrm{jet}}}\xspace}
\newcommand{\GeVc}{\ensuremath{\mathrm{GeV}/c}\xspace}
\newcommand{\pythia}{\textsc{Pythia}\xspace}
\newcommand{\haec}{HAEC\xspace}

\begin{document}

\title[Probing local charm-baryon formation with HAECs]{Probing local charm-baryon formation with hadron-anchored energy correlators}

\author{R\'obert V\'ertesi}

\address{HUN-REN Wigner Research Centre for Physics, Konkoly-Thege Mikl\'os \'ut 29-33, H-1121 Budapest, Hungary}
\ead{vertesi.robert@wigner.hu}

\begin{abstract}
We introduce the hadron-anchored energy correlator (HAEC), a jet-restricted realization of the fragmentation and semi-inclusive energy correlators, which measures the angular distribution of jet energy around a given hadron to access the local environment in which it is formed. We study the HAEC of \Dz- and \Lc-tagged jets in pp collisions at $\sqrt{s}=13$~TeV with \pythia~8, comparing hadronization without junctions to two color-reconnection scenarios with junction formation. The small-angle region of the correlator is dominated by strong-decay feed-down: the soft pion of $\Dstp\to\Dz\pi^+$ produces a kinematic edge at $\theta\,\langle\ptjet\rangle = 0.65~\GeVc/\langle\zh\rangle$, identical in all scenarios, which standard \Dstp reconstruction removes. The feed-down-cleaned correlator shows that at low momentum fraction \zh the energy around a \Lc is spread to larger angles than around a \Dz, an effect correlated with the displacement of the baryon from the jet axis and enhanced by junction formation. Counting (anti)protons around the anchor instead of weighting by energy isolates the compensation of the \Lc baryon number. Within an angular distance of 0.4 from the baryon, string-formed \Lc are accompanied by a net excess of 0.50 antiprotons per jet, peaked about 0.1~rad from the baryon and identical in all scenarios, whereas \Lc formed at a junction show less than half of this local compensation. The inclusive antiproton balance around \Lc therefore decreases linearly with the junction fraction, providing a measurable handle on how locally charm baryons are formed.
\end{abstract}

\noindent{\it Keywords\/}: hadronization, charm baryons, energy correlators, jets, baryon-number balance, color reconnection

\maketitle

\section{Introduction}
\label{sec:intro}

Charm-baryon production in proton--proton collisions is strongly enhanced relative to expectations from $\mathrm{e}^+\mathrm{e}^-$ annihilation. The $\Lc/\Dz$ ratio measured by ALICE is several times larger than in $\mathrm{e}^+\mathrm{e}^-$, decreases with transverse momentum, and depends on the event multiplicity~\cite{ALICE:2022exq,ALICE:2021npz,ALICE:2021rzj}. Similar enhancements are observed for $\Sigma_{\mathrm{c}}$, $\Xi_{\mathrm{c}}$ and $\Omega_{\mathrm{c}}$~\cite{ALICE:2021rzj,ALICE:2021bli,ALICE:2022cop}. The mechanism is not yet understood. Coalescence and statistical-hadronization approaches~\cite{Plumari:2017ntm,Minissale:2020bif,He:2019tik} and string fragmentation with color reconnection beyond leading color, which allows the formation of junctions~\cite{Christiansen:2015yqa}, all describe the enhanced yields, but they form the baryon in different ways. In \pythia, the enhancement has been linked to the underlying event and found to depend on the strangeness of the charm baryon~\cite{Varga:2021jzb,Varga:2023byp}. In ordinary string fragmentation a baryon is produced in a breakup with a diquark--antidiquark pair, so that its baryon number is compensated by an antibaryon produced nearby. At a junction, three string pieces meet and the baryon number is carried by the junction itself; the compensating antibaryon can be far away. In coalescence, the light (di)quarks are taken from a pre-existing partonic environment, so the antiquarks that balance their baryon number do not need to be close to the charm baryon. In statistical hadronization, baryon number is conserved globally or within a correlation volume that is generally much larger than a jet cone. Neither mechanism therefore implies the local compensation that is characteristic of string breaking, although quantitative predictions for jets require dynamical implementations of these models.

Existing measurements mostly characterize how many charm baryons are produced and how much momentum they carry, for example through the longitudinal momentum fraction $z_\parallel$ in jets~\cite{ALICE:2023jgm}. For \Dz mesons, the radial distribution with respect to the jet axis~\cite{CMS:2019jis,ALICE:2025ygv} and the angles between different jet-axis definitions~\cite{ALICE:2025ygv} have also been measured, probing where the charm hadron is placed relative to the jet. These observables do not, however, show how the remaining energy of the jet is arranged around the charm hadron itself, nor where the baryon number of a charm baryon is compensated. Energy correlators~\cite{Basham:1978bw,Chen:2020vvp,Komiske:2022enw} describe the angular organization of energy flow and have been applied to heavy-flavor jets to study the dead cone and the transition to hadrons~\cite{Craft:2022kdo,ALICE:2025igw,ALICE:2021aqk,ALICE:2022phr}. Two recent theoretical constructions anchor one leg of an energy correlator to an identified hadron. The fragmentation energy correlator (FEC)~\cite{Cao:2025icu} generalizes the fragmentation function by inserting an energy-flow operator into the same quark-field correlator, yielding an object differential in both the momentum fraction of the identified hadron and the angle of the surrounding radiation relative to it. The semi-inclusive energy correlator (SIEC)~\cite{Liu:2024kqt} extends the nucleon energy correlator~\cite{Liu:2022wop} to the fragmentation of a parton into an identified hadron, and has subsequently been studied in semi-inclusive $\mathrm{e}^+\mathrm{e}^-$ annihilation~\cite{Zhu:2025qkx}. Both are defined as fully inclusive quantities that correlate the identified hadron with the energy flow of the entire final state, and both give access to the energy environment of that hadron. In this paper we introduce a jet-restricted, discretized realization of the FEC and SIEC frameworks, the hadron-anchored energy correlator (\haec), which is directly measurable in heavy-flavor-tagged jets. Because the \haec is built from pairs of particles with one of them fixed to the identified hadron, the same construction can be resolved by the species of the partner particle. This allows us to follow the compensation of the baryon number of a charm baryon, which a one-particle observable such as the radial jet profile cannot do.

We define the \haec and its unweighted, baryon-number-resolved counterpart, together with the distance of the anchor from the jet axis as a complementary observable (section~\ref{sec:observables}), and study them for \Dz- and \Lc-tagged jets in \pythia~8 (section~\ref{sec:setup}). We show three things. First, the small-angle region of the correlator is dominated by strong-decay feed-down, which produces a sharp kinematic edge. We identify this edge quantitatively and show how to remove it with standard experimental techniques (section~\ref{sec:measuring}). Second, the cleaned correlator reveals a genuine baryon--meson difference: at low momentum fraction, the energy around a \Lc is spread to larger angles than around a \Dz, linked to the displacement of the baryon from the jet axis and enhanced by junction formation (section~\ref{sec:broadening}). Third, the baryon-number-resolved counterpart of the correlator locates the antibaryon that compensates the \Lc baryon number and separates string-formed from junction-formed charm baryons (section~\ref{sec:balance}). The inclusive antiproton balance around \Lc becomes a measurable discriminator of how locally charm baryons are formed (section~\ref{sec:discriminator}).

\section{Observables}
\label{sec:observables}

We define the \haec for a jet containing an identified hadron $h$ as the energy-weighted angular distribution of the other jet constituents around $h$,
\begin{equation}
  \mathcal{C}_h(\theta) = \frac{1}{N_\mathrm{jet}} \sum_\mathrm{jets} \sum_{i \neq h} z_h\, z_i\, \delta\big(\theta - \Delta R(h,i)\big),
  \label{eq:haec}
\end{equation}
with $z_h = p_\mathrm{T}^{h}/\ptjet$ and $z_i = p_{\mathrm{T},i}/\ptjet$, where $i$ runs over jet constituents, $\Delta R$ is the distance in the rapidity--azimuth plane and $N_\mathrm{jet}$ the number of $h$-tagged jets. Like the FEC~\cite{Cao:2025icu} and the SIEC~\cite{Liu:2024kqt,Zhu:2025qkx}, the \haec fixes one leg of the correlator to the identified hadron and is differential in its momentum fraction \zh. However, it is evaluated as a discrete sum over the final-state particles of the reconstructed jet containing $h$, rather than as an operator matrix element correlating $h$ with the energy flow of the full final state. This restriction connects the energy distribution directly to the jet environment in which the hadron is produced, as needed for heavy-flavor jet measurements. Species are compared through the ratio of the shape-normalized correlators,
\begin{equation}
  R_{\Lc/\Dz}(\theta) = \frac{\widehat{\mathcal{C}}_{\Lc}(\theta)}{\widehat{\mathcal{C}}_{\Dz}(\theta)},
  \label{eq:ratio}
  \qquad
  \widehat{\mathcal{C}}_h(\theta) = \frac{\mathcal{C}_h(\theta)}{\int_0^R \mathcal{C}_h(\theta')\,\mathrm{d}\theta'},  
\end{equation}
so that the ratio compares the shapes of the energy-weighted correlators of equation~\eqref{eq:haec}.
As a one-number summary of the shape we use the first moment of the correlator,
\begin{equation}
  \langle\theta\rangle_h = \frac{\int_0^R \theta\,\mathcal{C}_h(\theta)\,\mathrm{d}\theta}{\int_0^R \mathcal{C}_h(\theta)\,\mathrm{d}\theta},
  \label{eq:meanangle}
\end{equation}
which is the energy-weighted mean angular distance of the jet constituents from the anchor. The ratio $\langle\theta\rangle_{\Lc}/\langle\theta\rangle_{\Dz}$ then quantifies a shape difference between species in a single number.

Restricting the sum in equation~\eqref{eq:haec} to constituents of a given class $X$ and dropping the energy weights $z_h z_i$ gives the hadron-anchored pair density $\mathrm{d}N_\mathrm{pairs}^X/\mathrm{d}\theta$, the unweighted counterpart of the \haec. We use (anti)protons and classify them relative to the charm quantum number of the anchor: around a charm hadron (\Dz, \Lc), antiprotons carry baryon number \emph{opposite} (OB) to that of a \Lc, and protons the \emph{same} (SB); around anticharm hadrons the roles are reversed. The hadron-anchored baryon-number balance is the difference of the pair densities per jet,
\begin{equation}
  B_h(\theta) = \frac{1}{N_\mathrm{jet}}\left(\frac{\mathrm{d}N_\mathrm{pairs}^{\mathrm{OB}}}{\mathrm{d}\theta} - \frac{\mathrm{d}N_\mathrm{pairs}^{\mathrm{SB}}}{\mathrm{d}\theta}\right),
  \label{eq:balance}
\end{equation}
and its integral $\bar B_h = \int_0^R B_h(\theta)\,\mathrm{d}\theta$.
Only (anti)protons that are constituents of the jet containing $h$ enter these quantities, and $\bar B_h$ integrates over anchor--particle distances $\theta<R$. Since the anchor does not need to lie on the jet axis, a constituent can be up to $2R$ away from it. Therefore, $\bar B_h$ is a local balance around the anchor, and its relation to the balance of all jet constituents is discussed in section~\ref{sec:balance}. Baryon--antibaryon pairs produced independently of the anchor contribute equally to OB and SB and cancel, so $\bar B_h$ measures the net antiproton excess associated with the anchor, which in a given hadronization model can be related to the locality of the compensation of its baryon number. The construction follows the idea of balance functions~\cite{Bass:2000az}, which have been studied with \pythia for identified hadrons in inclusive pp collisions~\cite{Pruneau:2024jpa} and inside jets~\cite{Behera:2026bso}. The balance used here differs in that one particle of each pair is a reconstructed charm hadron, the partners are classified by baryon number relative to that anchor. In the simulation, the charm baryons can also be separated by their formation mechanism. The pair densities are unweighted because baryon number is carried by particles rather than by energy. However, the energy-weighted balance, obtained from equation~\eqref{eq:haec} restricted to the same classes, gives the same ordering of species and scenarios. Unlike equation~\eqref{eq:ratio}, the balance is normalized per jet and not to unit area, since the signal is the number of compensating antibaryons.

In addition, we use the distance $\Delta R(h,\hat n_\mathrm{jet})$ of the anchor from the jet axis, which characterizes where the hadron is placed in the jet. It has been measured for \Dz mesons~\cite{CMS:2019jis,ALICE:2025ygv}.

\section{Simulation setup}
\label{sec:setup}

Proton--proton collisions at $\sqrt{s}=13$~TeV are generated with \pythia~8.317~\cite{Sjostrand:2007gs,Bierlich:2022pfr} using hard $\mathrm{c}\bar{\mathrm{c}}$ production (\texttt{HardQCD:hardccbar}) with $\hat p_\mathrm{T}^\mathrm{min}=8~\GeVc$. \Dz and \Lc are kept stable; all strongly decaying states, including \Dstp, \Dstz, \Sc and \Scst, decay normally, and their decay products remain in the event record. Jets are reconstructed with the anti-$k_\mathrm{T}$ algorithm~\cite{Cacciari:2008gp} in FastJet~\cite{Cacciari:2011ma} with $R=0.4$ from charged final-state particles and the tagged \Dz or \Lc, which is included irrespective of its charge, mimicking the replacement of its decay daughters by the reconstructed hadron. Jets with $15<\ptjet<100~\GeVc$ are analyzed in four intervals (15--25, 25--40, 40--60, 60--100~\GeVc). If a jet contains more than one candidate, the one with the highest \pT is the anchor.

Three hadronization scenarios are compared (table~\ref{tab:settings}): the Monash 2013 tune~\cite{Skands:2014pea} with MPI-based color reconnection and no junctions; CR-QCD, the QCD-inspired reconnection model with junction formation~\cite{Christiansen:2015yqa} with all other parameters at their Monash values; and CR-BLC, the same reconnection model with its jointly retuned fragmentation parameters~\cite{Christiansen:2015yqa}. The samples contain $11.3\times10^6$, $8.7\times10^6$ and $8.2\times10^6$ \Dz-tagged jets and $0.80\times10^6$, $0.79\times10^6$ and $0.90\times10^6$ \Lc-tagged jets for Monash, CR-QCD and CR-BLC, respectively.

\begin{table}[t]
\caption{\pythia~8.317 settings of the three scenarios. Common to all: \texttt{Beams:eCM = 13000}, \texttt{Tune:pp = 14} (Monash 2013), \texttt{HardQCD:hardccbar = on}, \texttt{PhaseSpace:pTHatMin = 8}.}
\label{tab:settings}
\footnotesize
\centering
\begin{tabular}{@{}lccc@{}}
\br
Parameter & Monash & CR-QCD & CR-BLC \\
\mr
\texttt{ColourReconnection:reconnect}          & on & on & on \\
\texttt{ColourReconnection:mode}               & 0 & 1 & 1 \\
\texttt{ColourReconnection:allowJunctions}     & off & on & on \\
\texttt{ColourReconnection:allowDoubleJunRem}  & -- & on & off \\
\texttt{ColourReconnection:m0}                 & -- & 0.3 & 0.3 \\
\texttt{ColourReconnection:junctionCorrection} & -- & 1.20 & 1.20 \\
\texttt{ColourReconnection:timeDilationMode}   & -- & 2 & 2 \\
\texttt{ColourReconnection:timeDilationPar}    & -- & 0.18 & 0.18 \\
\texttt{StringPT:sigma}                        & 0.335 & 0.335 & 0.335 \\
\texttt{StringZ:aLund}                         & 0.68 & 0.68 & 0.36 \\
\texttt{StringZ:bLund}                         & 0.98 & 0.98 & 0.56 \\
\texttt{StringFlav:probQQtoQ}                  & 0.081 & 0.081 & 0.078 \\
\texttt{StringFlav:probStoUD}                  & 0.217 & 0.217 & 0.2 \\
\texttt{StringFlav:probQQ1toQQ0join}           & 0.5, 0.7, 0.9, 1.0 & 0.5, 0.7, 0.9, 1.0 & 0.0275 (all) \\
\texttt{MultipartonInteractions:pT0Ref}        & 2.28 & 2.28 & 2.15 \\
\texttt{BeamRemnants:remnantMode}              & 0 & 0 & 1 \\
\texttt{BeamRemnants:saturation}               & -- & -- & 5 \\
\br
\end{tabular}

\end{table}

To trace strong-decay feed-down, each anchor is classified by its direct mother: a \Dz from the decay of a \Dstp, from the decay of a \Dstz, or from neither of these, and a \Lc from the decay of a $\Sc(2455)$, from the decay of a $\Scst(2520)$, or from neither. Anchors in the last category are predominantly produced directly in the hadronization. The \emph{sisters} of an anchor are all other decay products of that mother. Only charged sisters enter the jets, so $\Dstz\to\Dz\pi^0/\gamma$ and $\Sc^+\to\Lc\pi^0$ do not contribute. Table~\ref{tab:feeddown} lists the feed-down fractions. The \Dstp fraction is the same in all scenarios, as it is set by the vector-to-pseudoscalar ratio, which is not varied. The \Sc fraction depends strongly on the scenario.

\begin{table}[t]
\caption{Feed-down fractions of the anchors (all jets, $15<\ptjet<100~\GeVc$).  The $\Sc(2455)$ and $\Scst(2520)$ columns include all three charge states. Only the neutral and doubly charged states decay with a charged pion.}
\label{tab:feeddown}
\footnotesize
\centering
\begin{tabular}{@{}lccccc@{}}
\br
& \multicolumn{2}{c}{\Dz from} & \multicolumn{3}{c}{\Lc from} \\
\cline{2-3}\cline{4-6}
Scenario & \Dstp & \Dstz & $\Sc(2455)$ & $\Scst(2520)$ & total \\
\mr
Monash & 25\% & 32\% & 12.3\% & 24.2\% & 36.5\% \\
CR-QCD & 25\% & 32\% & 21.4\% & 28.3\% & 49.7\% \\
CR-BLC & 25\% & 32\% & 40.1\% & 17.2\% & 57.3\% \\
\br
\end{tabular}

\end{table}

The \pythia status code of the primary hadron from which a \Lc descends (the \Lc itself, or the \Sc that decays into it) identifies how it was formed~\cite{Bierlich:2022pfr}: codes 83 and 84 denote normal string fragmentation (\emph{string}), 85 and 86 hadrons from the first two legs of a junction system (\emph{junction leg}), and 87--89 the baryon formed at the junction (\emph{junction baryon}). In Monash, 97.3\% of the \Lc are string-formed, 2.6\% come from junction legs and 0.1\% are junction baryons. Since the Monash scenario has no junction reconnection, these junctions arise from the beam remnants. In CR-QCD and CR-BLC, 63\% and 49\% of the \Lc are string-formed, 6\% and 5\% come from junction legs, while 31\% and 46\% are junction baryons, respectively.

All observables are also computed on the same events with an acceptance corresponding to the ALICE detector~\cite{ALICE:2008ngc,ALICE:2023udb}: $|\eta|<0.9$ for constituents and anchor, track $\pT>0.15~\GeVc$, and $|\eta_\mathrm{jet}|<0.9-R$, applied before clustering.

Statistical uncertainties of correlators and ratios are propagated from the bin errors. For the baryon balance they are estimated from the pair counts, treating pairs as independent. Since a jet can contribute more than one pair, this neglects correlations within jets and underestimates the true statistical uncertainty. However, with on average fewer than one (anti)proton per jet the effect is expected to be moderate. The differences between classes and scenarios discussed below exceed the quoted uncertainties by large factors, so this does not affect the conclusions.

\section{Strong-decay feed-down}
\label{sec:measuring}

\subsection{Small-angle structure of the raw correlator}

\begin{figure}[t]
  \centering
  \includegraphics[width=\textwidth]{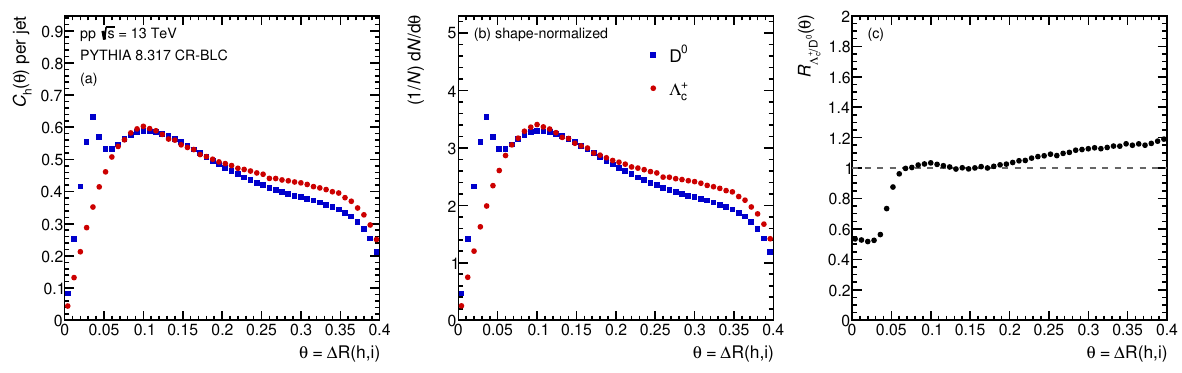}
  \caption{\haec of \Dz- and \Lc-tagged jets in the CR-BLC scenario, $15<\ptjet<100~\GeVc$: (a) per tagged jet, (b) normalized to unit area, (c) the ratio $R_{\Lc/\Dz}(\theta)$ of equation~\eqref{eq:ratio}.}
  \label{fig:baseline}
\end{figure}

Figure~\ref{fig:baseline} shows the \haec of \Dz- and \Lc-tagged jets and their ratio in the CR-BLC scenario. The ratio has a floor of $R_{\Lc/\Dz}\approx0.5$ at small angles, a sharp rise near $\theta\approx0.05$, and a slow increase towards 1.2 at the jet edge. A logistic fit
\begin{equation}
  R(\theta) = R_0 + \frac{R_\infty - R_0}{1+e^{-(\theta-\theta_0)/w}}
  \label{eq:sigmoid}
\end{equation}
in each jet-\pT interval (\ref{app:sigmoid}) gives a transition angle $\theta_0\propto1/\ptjet$ with $\theta_0\,\ptjet = 0.955\pm0.012$, $0.963\pm0.012$ and $0.946\pm0.014~\GeVc$ for Monash, CR-QCD and CR-BLC, with intercepts consistent with zero. The structure is therefore essentially independent of the hadronization model. It is strongest at high \zh: for $0.7<\zh<1$ the ratio dips to 0.35 below the transition and overshoots to about 1.55 just above it.

\subsection{Origin: decay sisters}

\begin{figure}[t]
  \centering
  \includegraphics[width=0.8\textwidth]{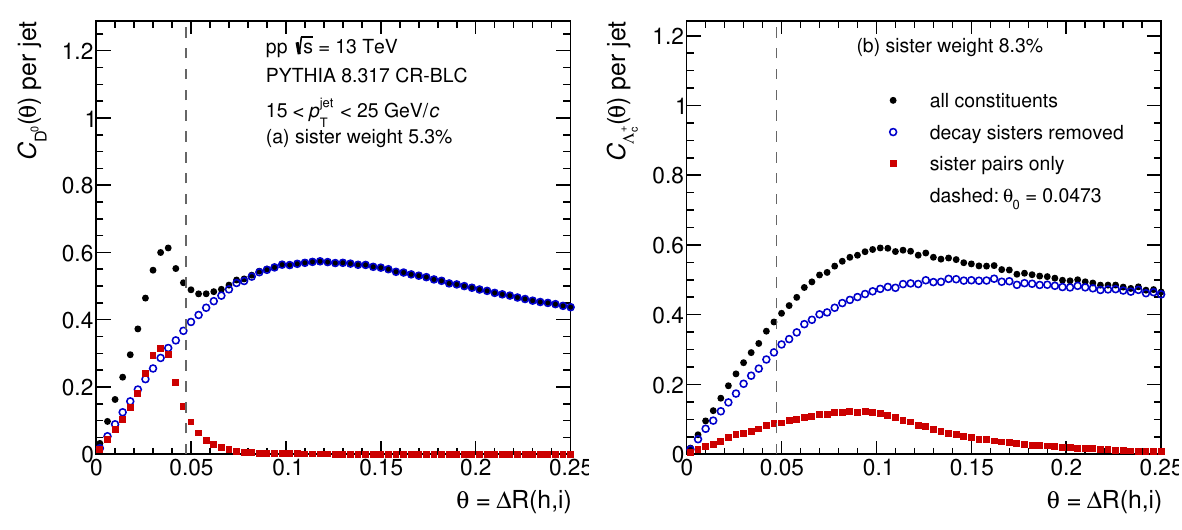}
  \caption{Decomposition of (a) the \Dz and (b) the \Lc correlator per jet at $15<\ptjet<25~\GeVc$ (CR-BLC): all constituents, decay sisters removed, and sister pairs only. The dashed line marks the fitted transition angle $\theta_0$ of the ratio.}
  \label{fig:decomposition}
\end{figure}

Separating the constituents into decay sisters of the anchor and all others identifies the origin (figure~\ref{fig:decomposition}). For \Dz-tagged jets, the soft pions from $\Dstp\to\Dz\pi^+$ form a distribution that rises towards a sharp edge and ends exactly at the fitted $\theta_0$; removing them leaves a smooth correlator. A quarter of all \Dz come from \Dstp, and their soft pion is always inside the same jet (table~\ref{tab:feeddown}). The sister pairs carry only 4--5\% of the total \Dz correlator weight (5.3\% at $15<\ptjet<25~\GeVc$, decreasing to 4.2\% at 60--100~\GeVc), but all of it lies below the edge, where the remaining correlator is small. For \Lc, the sisters from $\Sc^{(*)}\to\Lc\pi$ carry 6--8\% of the weight and are spread over a much wider angular range, peaking near $\theta\approx0.1$.

\subsection{The kinematic edge}

The edge follows from the kinematics of the two-body decay $M\to h\pi$, derived in \ref{app:kinematics}. For a highly boosted parent, an isotropic decay with rest-frame momentum $p^*$ and pion energy $E^*_\pi$ produces an opening-angle distribution between $h$ and $\pi$ that rises towards the edge
\begin{equation}
  \theta_\mathrm{max} \simeq \frac{M\,p^*}{E^*_\pi\,p_h},
  \label{eq:edge}
\end{equation}
where $p_h$ is the momentum of $h$. For $\Dstp\to\Dz\pi^+$ this gives $\theta_\mathrm{max}\,p_h\approx0.55~\GeVc$, while the exact maximum is $0.57~\GeVc$. For the \Sc decays the opening-angle distributions extend to considerably larger angles without a sharp edge. Considering that $p_{\mathrm{T},h}=\zh\,\ptjet$, in the rapidity--azimuth plane the \Dstp edge appears at $\Delta R\,\ptjet\simeq(M p^*/E^*_\pi)/\zh$.

\begin{figure}[t]
  \centering
  \includegraphics[width=0.85\textwidth]{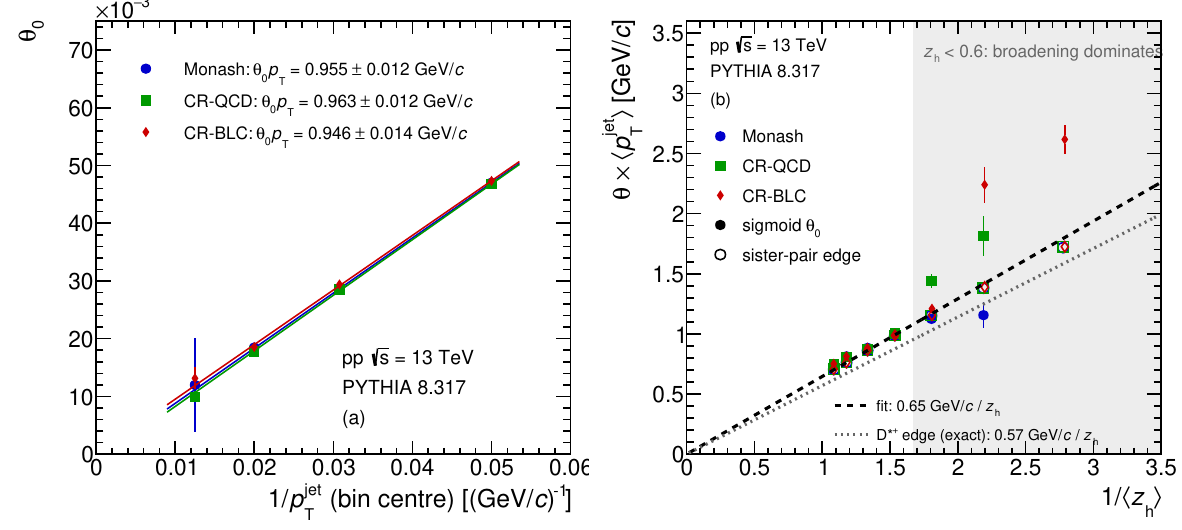}
  \caption{(a) Transition angle $\theta_0$ of the ratio versus $1/\ptjet$ for the three scenarios, with linear fits. (b) $\theta\,\langle\ptjet\rangle$ versus $1/\langle\zh\rangle$: sigmoid transition $\theta_0$ (filled) and edge of the \Dz sister-pair distribution (open), with a common fit $a/\zh$ to the sister edges (dashed) and the exact maximum of the $\Dstp\to\Dz\pi^+$ opening angle, $0.57~\GeVc/\zh$ (dotted). The shaded region marks $\zh<0.6$.}
  \label{fig:edge}
\end{figure}

Figure~\ref{fig:edge}(a) shows the $1/\ptjet$ scaling of the transition angle for all three scenarios, while figure~\ref{fig:edge}(b) tests the $1/\zh$ dependence. The edge of the \Dz sister-pair distribution, taken where it falls to half its maximum, follows $\theta\,\langle\ptjet\rangle = a/\langle\zh\rangle$ with $a=0.65~\GeVc$ in all three scenarios, which agree to the third decimal. 
Over $0.36<\langle\zh\rangle<0.92$ the measured edges deviate from this form by at most 4.3\%, and by less than 1\% for $\zh>0.6$, the range used in the fit.
A two-parameter fit $a/\zh+b$ gives $a=0.63~\GeVc$ and a negligible offset $b=0.025~\GeVc$. The coefficient is about 15\% above the exact maximum of $0.57~\GeVc$, consistent with the averaging over the falling jet-\pT and \zh distributions within each interval. For $\zh\ge0.6$, the sigmoid transition $\theta_0$ of the \Lc/\Dz ratio coincides with this edge to within one angular bin. At lower \zh the soft-pion weight, which scales as $\zh^2$, becomes small and the step of the ratio is masked by the genuine broadening discussed in section~\ref{sec:broadening}, while the sister edge itself continues to follow the kinematic scaling.

\subsection{Removing the feed-down}

\begin{figure}[t]
  \centering
  \includegraphics[width=0.85\textwidth]{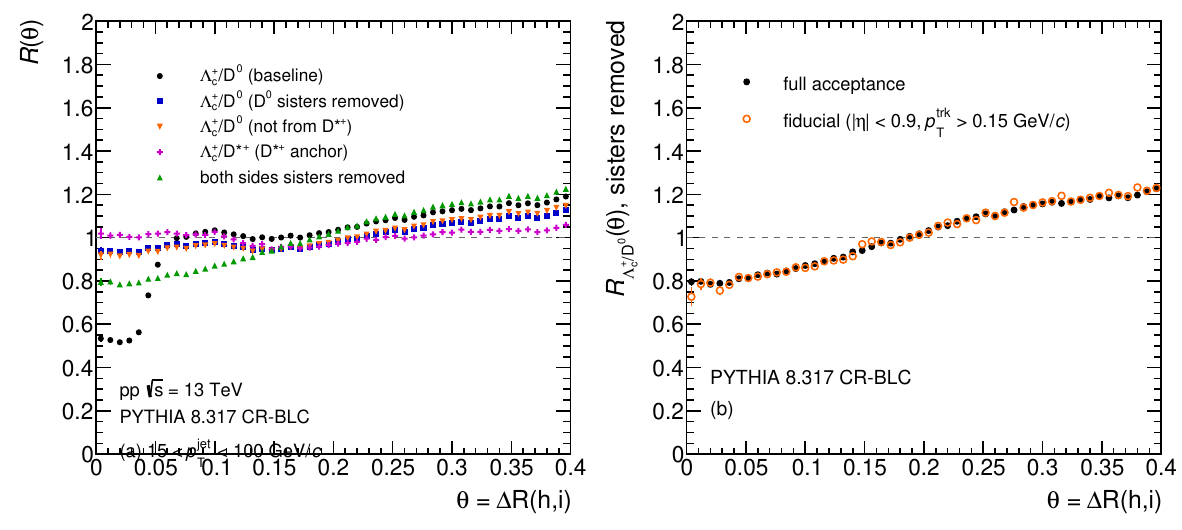}
  \caption{(a) $R(\theta)$ for different treatments of the feed-down (CR-BLC, $15<\ptjet<100~\GeVc$). (b) Ratio with sisters removed on both sides, full and fiducial acceptance.}
  \label{fig:cleaning}
\end{figure}

In simulations, three methods can be applied to remove the structure: excluding the sister pairs from the \Dz correlator, excluding \Dz from \Dstp altogether, and using the \Dstp as the anchor (see figure~\ref{fig:cleaning}(a)). The first two give identical ratios; with the \Dstp anchor the ratio stays close to unity up to $\theta\approx0.25$. Removing sisters on both sides yields a smooth, nearly linear rise from 0.8 to 1.2 with no localized transition. Experimentally, the first two options correspond to identifying the soft pion by reconstructing \Dstp candidates, either to exclude it from the correlator or to veto the \Dz, and the third to \Dstp-tagged jets. All three methods are standard in heavy-flavor analyses. They are demonstrated here at generator level and constitute a feasible strategy. However, their performance with a realistic detector response remains to be established. Removing the \Sc pions requires \Sc reconstruction, which is considerably harder; the half-cleaned ratio, with the \Dz cleaned and the \Lc inclusive, is therefore the most practical measurable version, and its \Lc correlator then also reflects the scenario-dependent \Sc feed-down fraction.

The mean-angle ratio $\langle\theta\rangle_{\Lc}/\langle\theta\rangle_{\Dz}$ is insensitive to the feed-down: removing the decay sisters changes it by less than 1\%. The soft pions distort the shape at small angles but hardly move the mean, so an imperfect removal of the feed-down in data has little effect on this quantity. With the fiducial acceptance, the cleaned ratio is unchanged within uncertainties and the corresponding mean-angle ratio changes from 1.072 to 1.075 (figure~\ref{fig:cleaning}(b)).

\section{Baryon--meson differences in the cleaned HAEC}
\label{sec:broadening}

\begin{figure}[t]
  \centering
  \includegraphics[width=\textwidth]{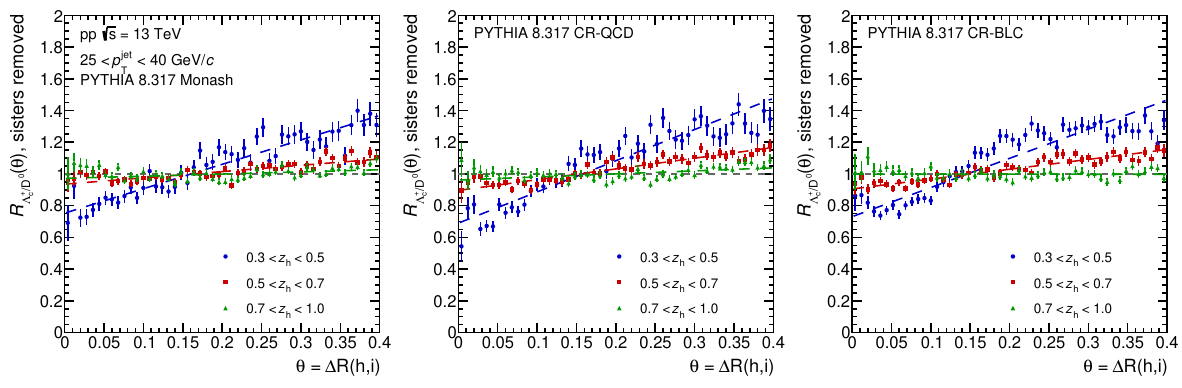}
  \caption{Ratio with decay sisters removed on both sides in three \zh intervals at $25<\ptjet<40~\GeVc$ for Monash, CR-QCD and CR-BLC, with straight-line fits.}
  \label{fig:broadening}
\end{figure}

With the feed-down removed on both sides, a genuine difference between \Lc and \Dz remains (figure~\ref{fig:broadening}). At low \zh, the energy around the \Lc is shifted to larger angles. The effect decreases with \zh and vanishes above $\zh\approx0.7$. Table~\ref{tab:slopes} lists the slopes of straight-line fits to the cleaned ratio at $25<\ptjet<40~\GeVc$. Junction formation enhances the broadening: CR-QCD, which adds only junctions to Monash, reproduces the CR-BLC slopes, which exceed the Monash values by 4--5 standard deviations in the two lower \zh intervals. The broadening is not merely compressed in angle at higher jet \pT, but its amplitude decreases as well, as expected for a soft, nonperturbative effect (\ref{app:scaling}).

\begin{table}[t]
\caption{Slopes (in rad$^{-1}$) of straight-line fits to the ratio with sisters removed on both sides, $0\le\theta\le0.4$, $25<\ptjet<40~\GeVc$.}
\label{tab:slopes}
\footnotesize
\centering
\begin{tabular}{@{}lccc@{}}
\br
\zh interval & Monash & CR-QCD & CR-BLC \\
\mr
0.3--0.5 & $1.54\pm0.07$ & $1.97\pm0.07$ & $1.84\pm0.05$ \\
0.5--0.7 & $0.40\pm0.04$ & $0.68\pm0.04$ & $0.62\pm0.04$ \\
0.7--1.0 & $0.14\pm0.03$ & $0.19\pm0.04$ & $0.00\pm0.04$ \\
\br
\end{tabular}

\end{table}

A difference in the \zh spectra within an interval could mimic a slope, since the \Lc spectrum is softer. In 0.05-wide intervals between $\zh=0.30$ and 0.50, however, the slope persists and decreases monotonically with \zh, from 2.8 to 1.3~rad$^{-1}$ in CR-BLC. Hence, the broadening is a property of the species at fixed \zh.

\begin{figure}[t]
  \centering
  \includegraphics[width=\textwidth]{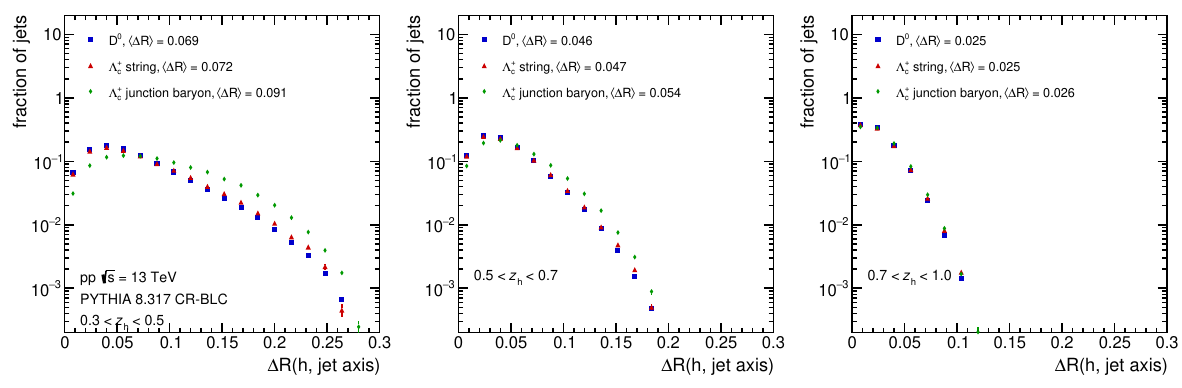}
  \caption{Distance of the anchor from the jet axis for \Dz, string-formed \Lc and junction-baryon \Lc in three \zh intervals (CR-BLC, $15<\ptjet<100~\GeVc$), normalized to unit area.}
  \label{fig:axis}
\end{figure}

The broadening is correlated with the position of the baryon in the jet (figure~\ref{fig:axis}, table~\ref{tab:axis}). At fixed \zh, string-formed \Lc are farther from the jet axis than \Dz, by about 15\% at $0.3<\zh<0.5$ in Monash and CR-QCD and by 5\% in CR-BLC. The \Lc formed at a junction are displaced further, by another 12--27\%. A displaced anchor sees energy centered on the jet axis at larger angles, which can account for part of the broadening. The two contributions are not separated here. At high \zh, where the anchor nearly coincides with the jet axis, all species agree. Note that at fixed \zh, the radial profile carries the same mean information as the \haec (\ref{app:profile}).

\begin{table}[t]
\caption{Mean anchor--axis distance $\langle\Delta R(h,\hat n_\mathrm{jet})\rangle$ in three \zh intervals ($15<\ptjet<100~\GeVc$). Statistical uncertainties are below $3\times10^{-4}$ except for the Monash junction-baryon class (820 jets; $\le0.005$).}
\label{tab:axis}
\footnotesize
\centering
\begin{tabular}{@{}llccc@{}}
\br
& & \multicolumn{3}{c}{\zh interval} \\
\cline{3-5}
Scenario & Anchor & 0.3--0.5 & 0.5--0.7 & 0.7--1.0 \\
\mr
\multirow{3}{*}{Monash} & \Dz & 0.0758 & 0.0477 & 0.0256 \\
 & \Lc string & 0.0868 & 0.0506 & 0.0256 \\
 & \Lc junc.\ baryon & 0.092 & 0.052 & 0.027 \\
\mr
\multirow{3}{*}{CR-QCD} & \Dz & 0.0736 & 0.0457 & 0.0247 \\
 & \Lc string & 0.0845 & 0.0479 & 0.0247 \\
 & \Lc junc.\ baryon & 0.0948 & 0.0551 & 0.0260 \\
\mr
\multirow{3}{*}{CR-BLC} & \Dz & 0.0686 & 0.0457 & 0.0245 \\
 & \Lc string & 0.0722 & 0.0466 & 0.0249 \\
 & \Lc junc.\ baryon & 0.0915 & 0.0543 & 0.0261 \\
\br
\end{tabular}

\end{table}

\section{Local baryon-number compensation}
\label{sec:balance}

\subsection{The compensating antibaryon}

\begin{figure}[t]
  \centering
  \includegraphics[width=0.85\textwidth]{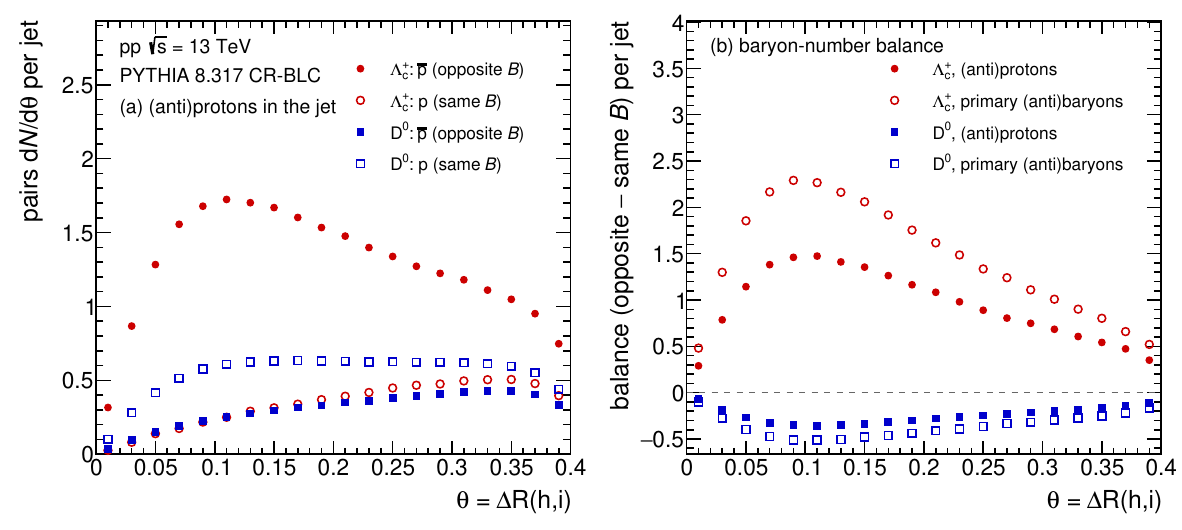}
  \caption{(a) Densities of antiprotons (opposite baryon number) and protons (same baryon number) around \Lc and \Dz per jet. (b) Balance $B_h(\theta)$ for (anti)protons (solid markers) and primary (anti)baryons (open markers). CR-BLC, $15<\ptjet<100~\GeVc$.}
  \label{fig:balance}
\end{figure}

Figure~\ref{fig:balance}(a) shows the proton and antiproton pair densities around \Lc and \Dz in CR-BLC. Around a \Lc, antiprotons are far more abundant than protons, while protons are more abundant around a \Dz. The balance $B_h(\theta)$ of \Lc (figure~\ref{fig:balance}(b)) peaks at $\theta\approx0.1$. Its mean angle is 0.176 for (anti)protons and 0.173 for all primary (anti)baryons (0.175--0.177 and 0.171--0.173 in the three scenarios, respectively). This indicates that restricting the measurement to final-state (anti)protons, which include the products of weak antihyperon decays, preserves the angular location of the compensating antibaryon. Integrated over $\theta<R$, the \Lc balance in CR-BLC is $\bar B=0.38$ for (anti)protons and $0.58$ for primary (anti)baryons (table~\ref{tab:balance}), which means that antiprotons retain about two thirds of the signal. Including the jet constituents beyond $\theta=R$ changes these values by 1--2\%. The negative \Dz balance reflects ordinary flavor flow in string fragmentation, where a proton can be formed next to a \Dz from the partner of its light antiquark. With the fiducial acceptance the \Lc balance decreases by 1--3\% and its mean angle by about 1\% in all three scenarios.

\subsection{String versus junction formation}

\begin{figure}[t]
  \centering
  \includegraphics[width=0.85\textwidth]{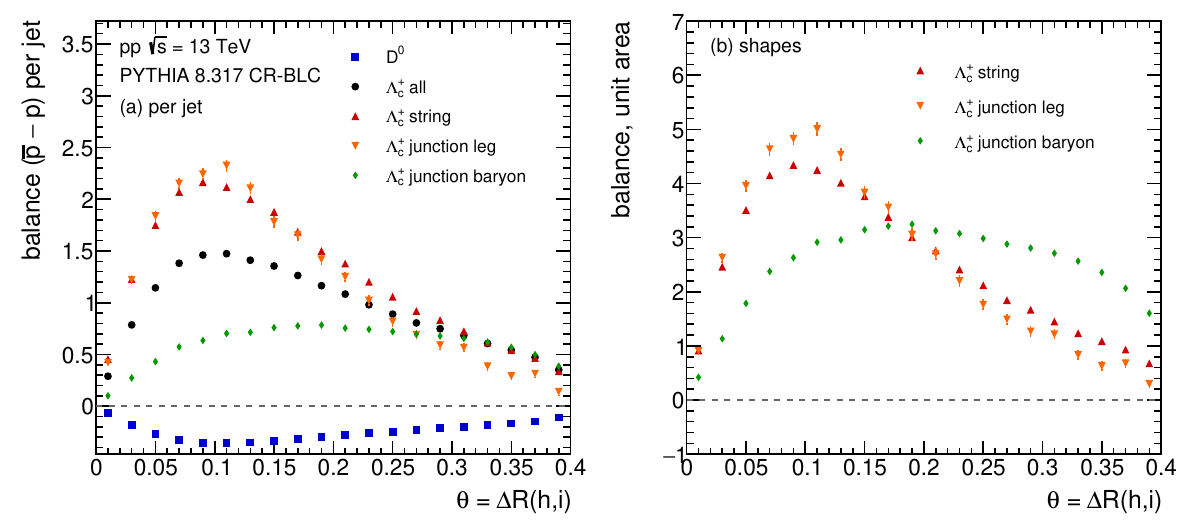}
  \caption{(a) Antiproton--proton balance per jet for \Dz, all \Lc, and \Lc of the three origin classes. (b) Shapes of the \Lc class balances, normalized to unit area. CR-BLC, $15<\ptjet<100~\GeVc$.}
  \label{fig:classes}
\end{figure}

Splitting the \Lc sample by origin (figure~\ref{fig:classes}, table~\ref{tab:balance}) shows that the local compensation depends on how the baryon was formed. The balance of string-formed \Lc is universal: $\bar B=0.499$, 0.495 and 0.499 in Monash, CR-QCD and CR-BLC, respectively. With $\bar B=0.75$--0.76 for primary (anti)baryons, the net antibaryon excess around a string-formed \Lc corresponds in \pythia to about three quarters of its baryon number within $\theta<R$. Hadrons from the legs of junction systems are formed by ordinary breakups and behave similarly, with $\bar B=0.46$--0.47. \Lc formed at a junction show a balance of only 0.21--0.24, less than half of the string value, with a broad, nearly flat angular distribution (figure~\ref{fig:classes}(b)): their baryon number is compensated locally to a much smaller extent. The few junction baryons in Monash, which arise from beam-remnant junctions, have $\bar B=0.03\pm0.02$, consistent with a baryon number supplied by the beam proton. We have checked that the class difference is not a \zh-composition effect: at fixed \zh the junction-baryon balance is 0.36--0.44 of the string value in every interval and scenario. Nor is it caused by the angular range of the integral. Although junction baryons are more displaced from the jet axis (section~\ref{sec:broadening}), including (anti)protons beyond $\theta=R$ raises the string balance by less than 1.2\% and the junction-baryon balance by 4--5\%. The junction-to-string ratio changes from 0.42 to 0.44 in CR-QCD and from 0.48 to 0.50 in CR-BLC, and by at most 0.02 in every \zh interval. The angular distribution itself reflects the difference in locality: only about 2\% of the antiprotons around a string-formed \Lc lie beyond $\theta=R$, compared with about 7\% around a junction baryon.

\begin{table}[t]
\caption{Junction-baryon fraction $f_\mathrm{jun}$ of \Lc and balance $\bar B$ (antiprotons minus protons per jet, $0\le\theta\le R$) for \Dz, all \Lc, and the \Lc origin classes; primary (anti)baryon balances in parentheses. Including the jet constituents beyond $\theta=R$ changes the magnitudes by at most 2\% for \Dz, all \Lc, string and junction-leg \Lc, and by about 5\% for junction baryons (section~\ref{sec:balance}). Statistical uncertainties are 0.001--0.002, except for the Monash junction-leg (0.006) and junction-baryon (0.016) classes.}
\label{tab:balance}
\footnotesize
\centering
\begin{tabular}{@{}lcccccc@{}}
\br
 & & & \multicolumn{4}{c}{\Lc} \\
\cline{4-7}
Scenario & $f_\mathrm{jun}$ & \Dz & all & string & junction leg & junction baryon \\
\mr
Monash & 0.001 & $-0.058$ ($-0.077$) & 0.498 (0.759) & 0.499 (0.761) & 0.472 (0.723) & 0.03 (0.06) \\
CR-QCD & 0.306 & $-0.094$ ($-0.136$) & 0.406 (0.626) & 0.495 (0.762) & 0.468 (0.694) & 0.209 (0.334) \\
CR-BLC & 0.463 & $-0.098$ ($-0.143$) & 0.378 (0.579) & 0.499 (0.753) & 0.464 (0.672) & 0.242 (0.385) \\
\br
\end{tabular}

\end{table}

\subsection{Momentum balance across the jet axis}

\begin{figure}[t]
  \centering
  \includegraphics[width=\textwidth]{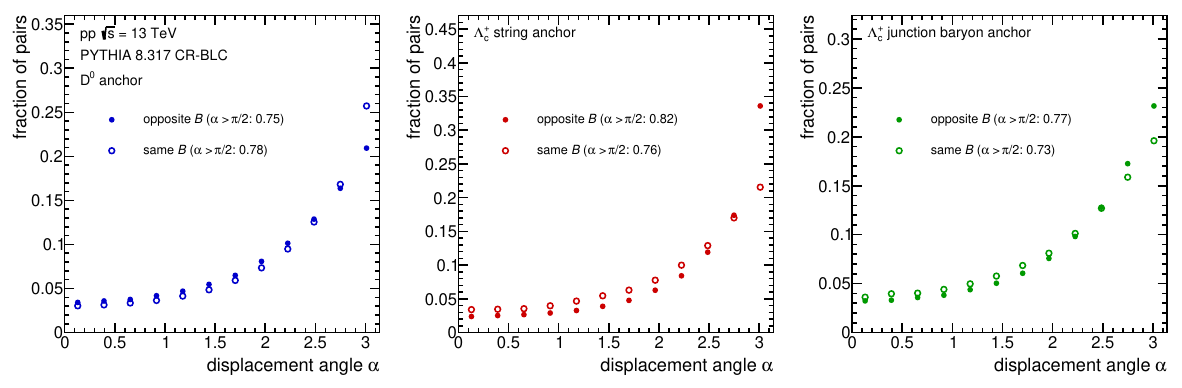}
  \caption{Distribution of the displacement angle $\alpha$ between the anchor and an (anti)proton, for opposite and same baryon number, around \Dz, string-formed \Lc and junction-baryon \Lc (CR-BLC), normalized to unit area. The fractions of pairs with $\alpha>\pi/2$, i.e.\ on opposite sides of the jet axis, are given in the legends.}
  \label{fig:azimuth}
\end{figure}

On which side of the jet axis the partner sits is measured by the displacement angle $\alpha$, the angle between the displacement vectors of the anchor and of an (anti)proton from the jet axis in the rapidity--azimuth plane, with $0\le\alpha\le\pi$ (figure~\ref{fig:azimuth}). Values of $\alpha$ near 0 correspond to the same side of the axis and values near $\pi$ to opposite sides; pairs in which either particle lies within $\Delta R=0.01$ of the axis, where the direction is undefined, are not included. Since the axis is the momentum sum of the constituents, any two particles tend to lie on opposite sides; the information is in the comparison. For string-formed \Lc, 82\% of the antiprotons but only 76\% of the protons have $\alpha>\pi/2$. For junction baryons the excess is smaller (77\% versus 73\%), and for \Dz it has the opposite sign (75\% versus 78\%). This supports transverse-momentum balance between a string-formed \Lc and its compensating antibaryon, which offers a natural explanation for the displacement of the baryon from the jet axis.

\section{Formation mechanism and experimental prospects}
\label{sec:discriminator}

\begin{figure}[t]
  \centering
  \includegraphics[width=\textwidth]{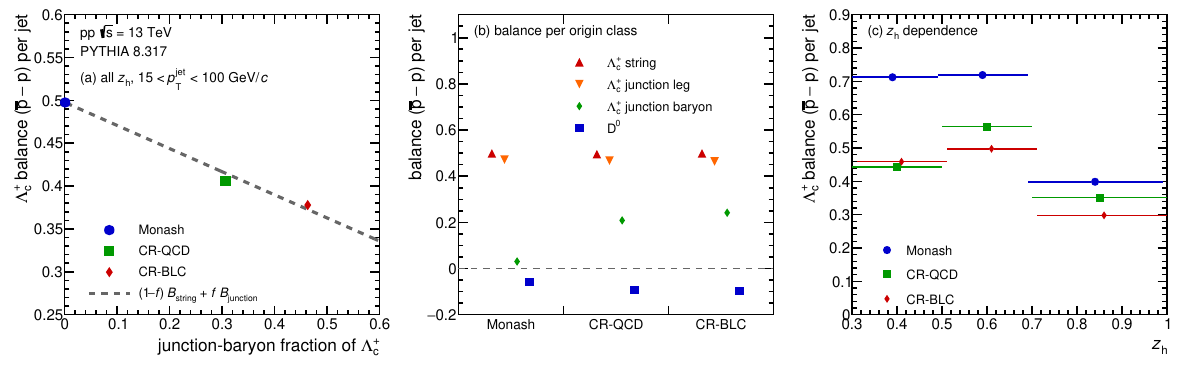}
  \caption{(a) Inclusive \Lc antiproton balance per jet versus the junction-baryon fraction for the three scenarios, with the two-component expectation (dashed). (b) Balance of the \Lc origin classes and of \Dz per scenario. The Monash junction baryons arise from beam-remnant junctions. (c) Inclusive \Lc balance in \zh intervals. $15<\ptjet<100~\GeVc$.}
  \label{fig:discriminator}
\end{figure}

Because the string balance is universal and the junction-baryon balance is much lower, the inclusive \Lc balance is determined by the fraction of \Lc formed at a junction (figure~\ref{fig:discriminator}(a)). The three scenarios, with junction fractions of 0.1\%, 31\% and 46\%, give $\bar B_{\Lc}=0.498$, 0.406 and 0.378 and lie within 0.01 on the two-component line $(1-f)\,B_\mathrm{string}+f\,B_\mathrm{junction}$ with $B_\mathrm{string}=0.498$ and $B_\mathrm{junction}=0.228$. The residual deviations reflect the small scenario dependence of the junction-baryon balance (figure~\ref{fig:discriminator}(b)). A linear dependence follows by construction from the two-component decomposition; the non-trivial content is that the string balance is universal and that the junction-baryon balance is much lower in both junction scenarios. The step from Monash to CR-QCD is a controlled test: CR-QCD adds junction formation to Monash and leaves all fragmentation parameters unchanged, and along this step the junction fraction rises from 0.1\% to 31\% and the inclusive balance falls from 0.498 to 0.406, while the string balance stays at 0.495--0.499. The relation between the inclusive balance and the junction fraction is nevertheless a model-specific calibration that demonstrates sensitivity to the formation mixture, but it cannot be transferred to data without a model.

The sensitivity is largest at low \zh (figure~\ref{fig:discriminator}(c), table~\ref{tab:balancez}): at $0.3<\zh<0.5$ the inclusive balances of Monash and CR-BLC differ by 0.25, compared with 0.10 at $0.7<\zh<1$. Junction baryons concentrate at low \zh, and the gap between the string and junction balances is largest there. In CR-BLC, 64\% of the \Lc at $0.3<\zh<0.5$ are junction baryons, compared with 36\% at $0.7<\zh<1$. The two-component decomposition reproduces these fractions from the balances alone: inverting $\bar B=(1-f)\,B_\mathrm{string}+f\,B_\mathrm{junction}$ with the class balances of table~\ref{tab:balancez} gives $f=0.64$ and 0.38, respectively. This is consistent with the concentration of the junction-induced enhancement of the $\Lc/\Dz$ ratio at low \pT~\cite{Christiansen:2015yqa,ALICE:2021rzj}.

\begin{table}[t]
\caption{Antiproton balance per jet in \zh intervals ($15<\ptjet<100~\GeVc$): all \Lc, string-formed \Lc, and junction-baryon \Lc. Statistical uncertainties are 0.001--0.006 (Monash junction baryons: 0.02--0.05).}
\label{tab:balancez}
\footnotesize
\centering
\begin{tabular}{@{}llccc@{}}
\br
& & \multicolumn{3}{c}{\zh interval} \\
\cline{3-5}
Scenario & \Lc & 0.3--0.5 & 0.5--0.7 & 0.7--1.0 \\
\mr
\multirow{3}{*}{Monash} & all & 0.712 & 0.719 & 0.399 \\
 & string & 0.716 & 0.721 & 0.400 \\
 & junc.\ baryon & 0.02 & 0.08 & $-0.01$ \\
\mr
\multirow{3}{*}{CR-QCD} & all & 0.443 & 0.564 & 0.351 \\
 & string & 0.706 & 0.717 & 0.411 \\
 & junc.\ baryon & 0.257 & 0.274 & 0.154 \\
\mr
\multirow{3}{*}{CR-BLC} & all & 0.459 & 0.498 & 0.298 \\
 & string & 0.730 & 0.684 & 0.378 \\
 & junc.\ baryon & 0.309 & 0.302 & 0.165 \\
\br
\end{tabular}

\end{table}

The interpretation in terms of a junction fraction is specific to \pythia. More generally, the balance measures the locality of baryon-number compensation: string breaking compensates the baryon number locally, junction formation does so to a much smaller extent, and neither coalescence nor statistical hadronization implies local compensation (see section~\ref{sec:intro}). Normalized to the universal string reference, the inclusive balance therefore measures how locally, on average, the baryon number of charm baryons is compensated compared with string fragmentation. Quantitative predictions for coalescence and statistical hadronization require dynamical implementations of these models that provide correlations at the jet level.

Experimentally, the balance requires identified (anti)protons in \Lc-tagged jets, which is standard in experiments with particle identification over the relevant momentum range. Since the balance is a difference of antiproton and proton yields, efficiency differences between them must be corrected. A \Dz-referenced version, $\bar B_{\Lc}-\bar B_{\Dz}$, further reduces detector effects. It preserves the ordering of the scenarios (0.556, 0.500 and 0.476), but the spread is reduced because the \Dz balance itself depends on the scenario. (Anti)protons from weak hyperon decays are included in the generator-level definition; experimentally they are partially rejected by track-selection criteria, which is accounted for by applying the same selection in the simulation. The proton from the decay $\Lc\to\mathrm{p}\mathrm{K}^-\pi^+$ is part of the reconstructed candidate and is excluded from the correlated particles, as in the simulation, where the \Lc is stable.

A simple estimate, treating the (anti)proton counts per jet as Poisson-distributed, gives $\sigma(\bar B)\simeq\sqrt{\langle n\rangle/(\varepsilon N)}$ for $N$ background-subtracted \Lc-tagged jets, with $\langle n\rangle\approx0.63$ (anti)protons per jet in the acceptance and an (anti)proton reconstruction and identification efficiency $\varepsilon$. Distinguishing the Monash and CR-BLC balances, which differ by 0.12 (table~\ref{tab:balance}), at the level of five standard deviations then requires $N\approx1.1\times10^3/\varepsilon$. Note that the present study covers $15<\ptjet<100~\GeVc$, above the range of the existing measurement of \Lc-tagged jets~\cite{ALICE:2023jgm}. Establishing the experimental reach requires extending the simulation to lower jet \pT.

\section{Conclusions and outlook}
\label{sec:conclusions}

We have introduced the hadron-anchored energy correlator, a jet-restricted realization of the FEC and SIEC frameworks, together with a hadron-anchored baryon-number balance built from the same anchored pairs. We have shown that it gives access to the local environment in which a heavy-flavor hadron is formed. In \Dz- and \Lc-tagged jets simulated with \pythia~8, the small-angle region is dominated by strong-decay feed-down: the \Dstp soft pion produces a kinematic edge at $\theta\,\langle\ptjet\rangle=0.65~\GeVc/\langle\zh\rangle$, identical in all hadronization scenarios, which standard \Dstp reconstruction removes. The mean angle of the correlator is insensitive to the feed-down, and the cleaned correlator is unaffected by an ALICE-like acceptance. The cleaned correlator shows that at low \zh the energy around a \Lc is spread to larger angles than around a \Dz, correlated with the displacement of the baryon from the jet axis and enhanced by junction formation. Resolving the correlated particles by baryon number locates the antibaryon that compensates the \Lc. String-formed \Lc correspond to a universal net antiproton excess of 0.50 per jet within an angular distance of 0.4, while junction-formed \Lc show less than half of it. Together, the energy flow and the baryon-number balance give a consistent picture, in which string-formed \Lc recoil against a nearby compensating antibaryon, while \Lc formed at a junction, which are compensated less locally, are displaced further from the jet axis. The inclusive antiproton balance around \Lc therefore decreases linearly with the junction fraction and provides a measurable handle on how locally charm baryons are formed.

Several extensions follow directly. A second hadronization paradigm, such as cluster hadronization or coalescence, will test whether the locality of compensation separates hadronization mechanisms beyond \pythia. In the beauty sector, $\mathrm{B}^*\to\mathrm{B}\gamma$ produces no charged soft pion, so the meson reference is free of the feed-down edge, while the pions from $\Sigma_\mathrm{b}^{(*)}\to\Lb\pi$ extend to $\theta\,p_h$ of about 5.4--6.5~\GeVc (exact maxima, see \ref{app:kinematics}), compared with 1.6--3.2~\GeVc for $\Sc^{(*)}\to\Lc\pi$ and 0.57~\GeVc for the \Dstp edge. For strange charm baryons, the compensation of strangeness can be traced with identified kaons in the same way. Finally, hadron-anchored correlators of light-flavor hadrons with large strong-decay feed-down fractions are expected to show analogous kinematic edges in the region of the hadronization transition.

\ack
This work has been supported by the Hungarian National Research, Development and Innovation Office (NKFIH) under contract numbers NKKP ADVANCED 25-153456 and 2025-1.1.5-NEMZ KI-2025-00002, the Wigner Scientific Computing Laboratory (WSCLAB), and the HUN-REN Cloud.
During the preparation of this work, the author used Claude Opus 5.5 (Anthropic) to assist with code development including the analysis and plotting code used to produce the figures, with physics analysis (in particular, the derivation presented in \ref{app:kinematics}) and with editing the text and formatting the tables. The author specified all the analyses, reviewed, validated and critically revised all AI-assisted outputs and takes full responsibility for the content of the work.

\appendix

\section{Sigmoid fits to the raw ratio}
\label{app:sigmoid}

\begin{figure}[t]
  \centering
  \includegraphics[width=0.7\textwidth]{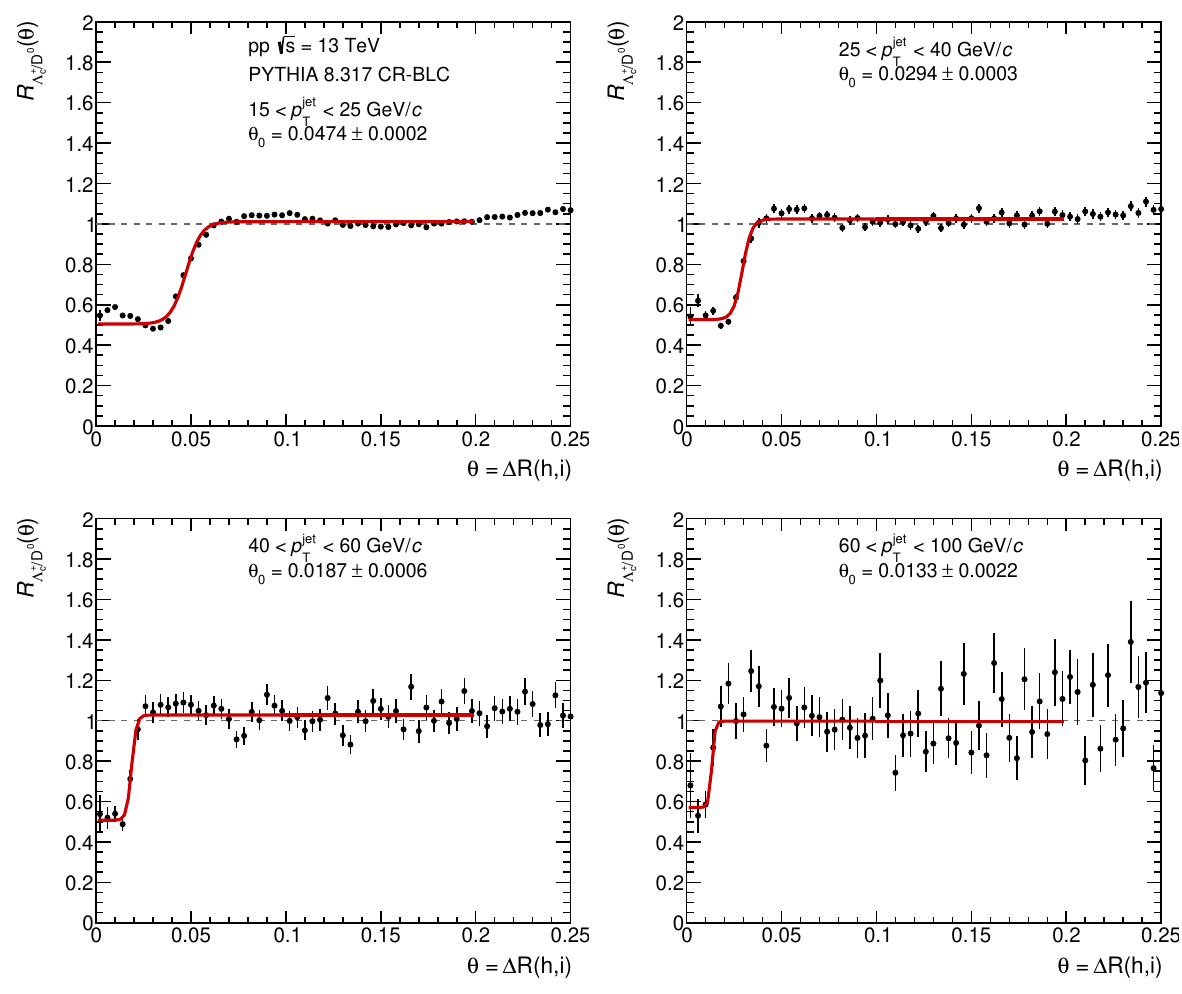}
  \caption{Logistic fits (equation~\eqref{eq:sigmoid}) to the raw $R_{\Lc/\Dz}(\theta)$ in four jet-\pT intervals (CR-BLC).}
  \label{fig:sigmoid}
\end{figure}

Figure~\ref{fig:sigmoid} shows the fits of equation~\eqref{eq:sigmoid} to the raw ratio in the four jet-\pT intervals for CR-BLC. The fitted transition angles are $\theta_0=0.0473$, 0.0294, 0.0186 and 0.0131 for increasing jet \pT.

\section{Kinematics of the decay edge}
\label{app:kinematics}

Consider a parent of mass $M$ moving with laboratory momentum $P$ along the $z$ axis, with Lorentz factor $\gamma$ and $\gamma\beta=P/M$, that decays into a hadron $h$ and a pion. In the parent rest frame both daughters have momentum $p^*$, the pion has energy $E^*_\pi$ and the hadron $E^*_h=M-E^*_\pi$, and the pion is emitted at the angle $\theta^*$ with respect to the $z$ axis. In the laboratory both daughters carry the transverse momentum $p^*\sin\theta^*$, in opposite directions, and the longitudinal momenta
\begin{align}
  p_{\parallel,\pi} &= \gamma\left(\beta E^*_\pi + p^*\cos\theta^*\right),\\
  p_{\parallel,h} &= \gamma\left(\beta E^*_h - p^*\cos\theta^*\right).
\end{align}
For a highly boosted parent ($\beta\to1$) both daughters are emitted at small angles on opposite sides of the parent direction, and their opening angle is
\begin{equation}
  \theta_{h\pi} \simeq \frac{p^*\sin\theta^*}{\gamma}\left(\frac{1}{E^*_\pi+p^*\cos\theta^*}+\frac{1}{E^*_h-p^*\cos\theta^*}\right).
  \label{eq:openexact}
\end{equation}

To leading order in $p^*/E^*_\pi$ the $\cos\theta^*$ terms can be neglected. Using $E^*_\pi+E^*_h=M$ and expressing the boost through the momentum of the hadron, $p_h\simeq\gamma E^*_h$, gives
\begin{equation}
  \theta_{h\pi} \simeq \frac{M\,p^*\sin\theta^*}{E^*_\pi\,p_h},
  \label{eq:openlo}
\end{equation}
which is maximal for $\theta^*=\pi/2$ and yields equation~\eqref{eq:edge}. For an isotropic decay $\cos\theta^*$ is uniformly distributed, so $x=\sin\theta^*$ follows $\mathrm{d}N/\mathrm{d}x\propto x/\sqrt{1-x^2}$: the opening-angle distribution rises towards the edge and ends there.

Beyond leading order, the pion term dominates equation~\eqref{eq:openexact}, since $E^*_\pi\ll E^*_h$. For slightly backward emission, $\cos\theta^*<0$, the pion is slower in the laboratory while its transverse momentum is nearly unchanged, so the opening angle grows. The function $\sin\theta^*/(E^*_\pi+p^*\cos\theta^*)$ is maximal at $\cos\theta^*=-p^*/E^*_\pi$, where it equals $1/m_\pi$. The exact maximum is therefore well approximated by
\begin{equation}
  \theta_\mathrm{max}\,p_h \simeq \frac{M\,p^*}{m_\pi},
  \label{eq:edgeexact}
\end{equation}
which exceeds the leading-order value by the factor $E^*_\pi/m_\pi$. At small angles the relative transverse momentum of the two daughters is $k_\perp\simeq\theta\,p\simeq\Delta R\,p_\mathrm{T}$, so the same edge appears in the rapidity--azimuth plane as a maximal $\Delta R\,p_{\mathrm{T},h}$.

Table~\ref{tab:edges} lists the values for the decays discussed in this paper, using standard masses~\cite{ParticleDataGroup:2024cfk}; the exact maxima are obtained by maximizing $\theta_{h\pi}\,p_h$ numerically over $\theta^*$ with the full two-body kinematics. For $\Dstp\to\Dz\pi^+$, where $p^*/E^*_\pi=0.27$, the leading-order value is accurate to 4\%. For the \Sc and $\Sigma_\mathrm{b}$ decays, $p^*/E^*_\pi=0.5$--0.8, and the exact maxima, reached by slow, backward-emitted pions, exceed the leading-order values substantially. These pions carry little weight in the correlator, so the distributions show no sharp edge.

\begin{table}[t]
\caption{Maximal opening angle between the daughters of strong two-body decays, $\theta_\mathrm{max}\,p_h$ in \GeVc: leading order (LO) (equation~\eqref{eq:edge}), the approximation of equation~\eqref{eq:edgeexact}, and the exact maximum.}
\label{tab:edges}
\footnotesize
\centering
\begin{tabular}{@{}lcccc@{}}
\br
Decay & $p^*$ (MeV/$c$) & LO & $M p^*/m_\pi$ & Exact \\
\mr
$\Dstp\to\Dz\pi^+$ & 39 & 0.55 & 0.57 & 0.57 \\
$\Sc(2455)\to\Lc\pi$ & 89 & 1.32 & 1.57 & 1.57 \\
$\Scst(2520)\to\Lc\pi$ & 177 & 1.98 & 3.19 & 3.19 \\
$\Sigma_\mathrm{b}\to\Lb\pi$ & 128 & 3.93 & 5.34 & 5.37 \\
$\Sigma_\mathrm{b}^*\to\Lb\pi$ & 155 & 4.33 & 6.47 & 6.53 \\
\br
\end{tabular}

\end{table}

\section{Jet-\texorpdfstring{$p_\mathrm{T}$}{pT} dependence of the cleaned ratio}
\label{app:scaling}

\begin{figure}[t]
  \centering
  \includegraphics[width=\textwidth]{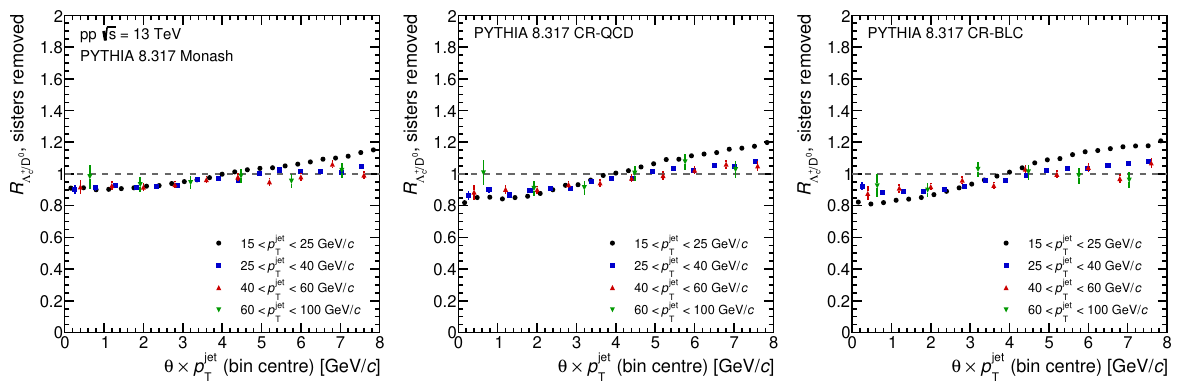}
  \caption{Ratio with decay sisters removed on both sides versus $\theta\,\ptjet$ (bin center) in four jet-\pT intervals.}
  \label{fig:scaling}
\end{figure}

Plotted against $\theta\,\ptjet$ (figure~\ref{fig:scaling}), the cleaned ratios of the four jet-\pT intervals do not collapse onto one curve. In Monash they agree within uncertainties up to $\theta\,\ptjet\approx4~\GeVc$; in CR-QCD and CR-BLC the lowest interval lies below the others at small $\theta\,\ptjet$. The \zh composition of the samples changes with jet \pT, so a scaling test at fixed \zh requires more statistics at high \pT.

\section{Comparison with the radial profile}
\label{app:profile}

\begin{figure}[t]
  \centering
  \includegraphics[width=\textwidth]{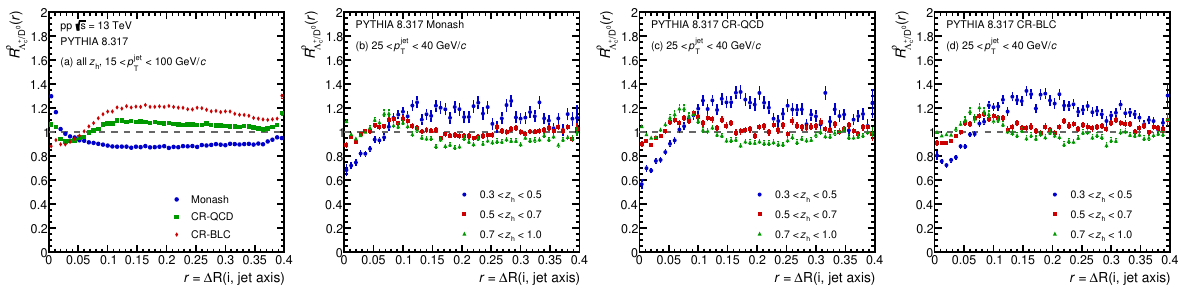}
  \caption{Ratio of radial profiles $\rho(r)$ of \Lc- and \Dz-tagged jets: (a) all \zh, three scenarios; (b)--(d) three \zh intervals at $25<\ptjet<40~\GeVc$.}
  \label{fig:profile}
\end{figure}

For comparison we also use the conventional radial momentum profile
\begin{equation}
  \rho(r) = \frac{1}{N_\mathrm{jet}} \sum_\mathrm{jets} \sum_{i} z_i\, \delta\big(r - \Delta R(i,\hat n_\mathrm{jet})\big),
  \label{eq:profile}
\end{equation}
which includes the anchor and is referenced to the jet axis rather than to a physical particle. The inclusive radial-profile ratio (figure~\ref{fig:profile}(a)) lies below unity in Monash and above unity in CR-BLC, with CR-QCD in between, a much larger spread than for the \haec. In \zh intervals (figure~\ref{fig:profile}(b)--(d)) the profile and the \haec carry the same mean information: their mean-distance ratios agree within 0.6\% for $\zh<0.7$ (table~\ref{tab:means}). The inclusive difference is a \zh-composition effect: the profile includes the anchor with weight \zh and is therefore sensitive to the different \zh spectra of the scenarios.

The two observables are therefore complementary rather than one superseding the other. The radial profile is nearly insensitive to strong-decay feed-down, since a soft decay pion lies at almost the same distance from the jet axis as its parent, and as a one-particle observable it is simpler to measure. At fixed \zh it carries the same first-moment information as the \haec, while inclusively it is dominated by the \zh spectra, which are also accessible through measurements of the longitudinal momentum fraction~\cite{ALICE:2023jgm}. The \haec instead excludes the anchor and is referenced to it rather than to an algorithmically defined axis; together with the anchor--axis distance, it separates where the hadron is placed in the jet from how the remaining energy is arranged around it. Its decisive advantage is its pair structure: correlating the anchor with identified particles yields the hadron-anchored baryon-number balance of section~\ref{sec:balance}, which has no counterpart in the radial profile. Finally, as a jet-level realization of the FEC and SIEC, the \haec connects to field-theoretic definitions that open a route to calculations.

\begin{table}[t]
\caption{\Lc/\Dz ratio of mean distances: radial profile $\langle r\rangle$ and \haec $\langle\theta\rangle$ with sisters removed. Inclusive rows: all \zh, $15<\ptjet<100~\GeVc$ (uncertainties $<0.001$); \zh intervals: $25<\ptjet<40~\GeVc$ (uncertainties 0.002--0.005).}
\label{tab:means}
\footnotesize
\centering
\begin{tabular}{@{}llcc@{}}
\br
Scenario & \zh & profile $\langle r\rangle$ & \haec $\langle\theta\rangle$ \\
\mr
\multirow{4}{*}{Monash} & all & 0.913 & 1.037 \\
 & 0.3--0.5 & 1.103 & 1.104 \\
 & 0.5--0.7 & 1.015 & 1.023 \\
 & 0.7--1.0 & 0.969 & 1.005 \\
\mr
\multirow{4}{*}{CR-QCD} & all & 1.031 & 1.061 \\
 & 0.3--0.5 & 1.130 & 1.135 \\
 & 0.5--0.7 & 1.039 & 1.044 \\
 & 0.7--1.0 & 0.986 & 1.008 \\
\mr
\multirow{4}{*}{CR-BLC} & all & 1.113 & 1.072 \\
 & 0.3--0.5 & 1.132 & 1.132 \\
 & 0.5--0.7 & 1.043 & 1.042 \\
 & 0.7--1.0 & 0.998 & 0.998 \\
\br
\end{tabular}

\end{table}

\section*{References}
\bibliographystyle{iopart-num-long}
\bibliography{haec_paper}

\end{document}